\documentclass[sigconf]{acmart}
\renewcommand\footnotetextcopyrightpermission[1]{}

\usepackage{url}  %Required
\usepackage{graphicx}  %Required
\usepackage{amsfonts}
\usepackage{amssymb,amsmath}
\usepackage{xspace}
\usepackage{booktabs}
\usepackage{amsmath,bm}
\usepackage{dsfont}
\usepackage{multirow}
\usepackage [autostyle, english = american]{csquotes}
\usepackage{array}
\newcolumntype{L}[1]{>{\raggedright\let\newline\\\arraybackslash\hspace{0pt}}m{#1}}
\newcolumntype{C}[1]{>{\centering\let\newline\\\arraybackslash\hspace{0pt}}m{#1}}
\newcolumntype{R}[1]{>{\raggedleft\let\newline\\\arraybackslash\hspace{0pt}}m{#1}}

\usepackage{algorithm}% http://ctan.org/pkg/algorithm
\PassOptionsToPackage{noend}{algpseudocode}% comment out if want end's to show
\usepackage{algpseudocode}% http://ctan.org/pkg/algorithmicx

\usepackage{etoolbox}
\usepackage{tikz}
\usetikzlibrary{tikzmark}
\usetikzlibrary{calc}
\algblock[TryCatchFinally]{try}{endtry}
\algcblock[TryCatchFinally]{TryCatchFinally}{finally}{endtry}
\algcblockdefx[TryCatchFinally]{TryCatchFinally}{catch}{endtry}
	[1]{\textbf{catch} #1}
	{\textbf{end try}}
\newcommand{\ALGtikzmarkcolor}{black}% customise this, if you want
\newcommand{\ALGtikzmarkextraindent}{4pt}% customise this, if you want
\newcommand{\ALGtikzmarkverticaloffsetstart}{-.5ex}% customise this, if you want
\newcommand{\ALGtikzmarkverticaloffsetend}{-.5ex}% customise this, if you want
\makeatletter
\newcounter{ALG@tikzmark@tempcnta}
\newcommand\ALG@tikzmark@start{%
    \global\let\ALG@tikzmark@last\ALG@tikzmark@starttext%
    \expandafter\edef\csname ALG@tikzmark@\theALG@nested\endcsname{\theALG@tikzmark@tempcnta}%
    \tikzmark{ALG@tikzmark@start@\csname ALG@tikzmark@\theALG@nested\endcsname}%
    \addtocounter{ALG@tikzmark@tempcnta}{1}%
}
\def\ALG@tikzmark@starttext{start}
\newcommand\ALG@tikzmark@end{%
    \ifx\ALG@tikzmark@last\ALG@tikzmark@starttext
    \else
        \tikzmark{ALG@tikzmark@end@\csname ALG@tikzmark@\theALG@nested\endcsname}%
        \tikz[overlay,remember picture] \draw[\ALGtikzmarkcolor] let \p{S}=($(pic cs:ALG@tikzmark@start@\csname ALG@tikzmark@\theALG@nested\endcsname)+(\ALGtikzmarkextraindent,\ALGtikzmarkverticaloffsetstart)$), \p{E}=($(pic cs:ALG@tikzmark@end@\csname ALG@tikzmark@\theALG@nested\endcsname)+(\ALGtikzmarkextraindent,\ALGtikzmarkverticaloffsetend)$) in (\x{S},\y{S})--(\x{S},\y{E});%
    \fi
    \gdef\ALG@tikzmark@last{end}%
}
\apptocmd{\ALG@beginblock}{\ALG@tikzmark@start}{}{\errmessage{failed to patch}}
\pretocmd{\ALG@endblock}{\ALG@tikzmark@end}{}{\errmessage{failed to patch}}
\makeatother
\newcommand{\mname}{\texttt{CardioLearn}\xspace}
\usepackage[utf8]{inputenc}
\usepackage{color}

\begin{document}
\fancyhead{}

\title{\mname: A Cloud Deep Learning Service for Cardiac Disease Detection from Electrocardiogram}

\author{Shenda~Hong$^{1}$, Zhaoji~Fu$^{1,2}$, Rongbo~Zhou$^{1}$, Jie~Yu$^{1}$, Yongkui~Li$^1$, Kai~Wang$^1$, Guanlin~Cheng$^1$}
\affiliation{$^1$HeartVoice Medical Technology, Hefei, China\\ $^2$University of Science and Technology of China, Hefei, China}

\begin{abstract}

Electrocardiogram (ECG) is one of the most convenient and non-invasive tools for monitoring peoples' heart condition, which can use for diagnosing a wide range of heart diseases, including Cardiac Arrhythmia, Acute Coronary Syndrome, et al. However, traditional ECG disease detection models show substantial rates of misdiagnosis due to the limitations of the abilities of extracted features. Recent deep learning methods have shown significant advantages, but they do not provide publicly available services for those who have no training data or computational resources. 

In this paper, we demonstrate our work on building, training, and serving such out-of-the-box cloud deep learning service for cardiac disease detection from ECG named \mname. The analytic ability of any other ECG recording devices can be enhanced by connecting to the Internet and invoke our open API. As a practical example, we also design a portable smart hardware device along with an interactive mobile program, which can collect ECG and detect potential cardiac diseases anytime and anywhere.

\end{abstract}


\maketitle

\section{Introduction}

The Electrocardiogram (ECG) is one of the most convenient and non-invasive tools for monitoring peoples' heart condition. It is a kind of physiological signal that records electrical activities of cardiac muscle over a period of time by placing electrodes and leads on the human body. 
ECG can use for diagnosing a wide range of heart diseases, including Cardiac Arrhythmia, Acute Coronary Syndrome, et al \cite{yanowitz2012introduction,garcia2014introduction}. It is estimated that more than 300 million ECGs are recorded worldwide every year \cite{hannun2019cardiologist}, which is a tremendous amount of data for Cardiologists to analyze. Thus, many computer-aided ECG disease detection methods based on feature extraction and machine learning have been proposed over the past 50 years, and they have been used in commercial medical devices. However, existing commercial ECG disease detection methods still show substantial rates of misdiagnosis \cite{schlapfer2017computer,shah2007errors,guglin2006common}, due to the limitations of the abilities of extracted features, and the lack of generalizability which are tuned for their specific medical devices. 

Recently, deep learning methods have shown great potential in healthcare and medical area \cite{miotto2017deep,xiao2018opportunities}. Specifically, there are some pioneer works that show successes of deep learning methods on ECG disease detection \cite{hannun2019cardiologist,attia2019screening,hong2019mina,zhou2019k,xu2018raim,shashikumar2018detection,hong2017encase} (see \cite{hong2019opportunities} for a survey). However, these methods are still far away from practical applications because none of these models have been deployed for providing publicly available ECG disease detection services. Besides, most of these models only trained on single lead \footnote{Here ``lead'' means ``channel''.} ECG data thus, they only support single lead ECG disease detection, which is insufficient in most medical area applications.

This demonstration provides \mname, a publicly available out-of-the-box cloud deep learning service that can be used for cardiac disease detection from ECG. Any existing ECG recording devices can be enhanced with cardiac disease detection ability by connecting to the Internet and invoke our open API. To further demonstrate such practical usage, we also design a portable smart hardware device along with an interactive mobile program, so that people can easily collect their ECG records and detect potential cardiac diseases anytime and anywhere.

\begin{figure}
\includegraphics[width=0.45\textwidth]{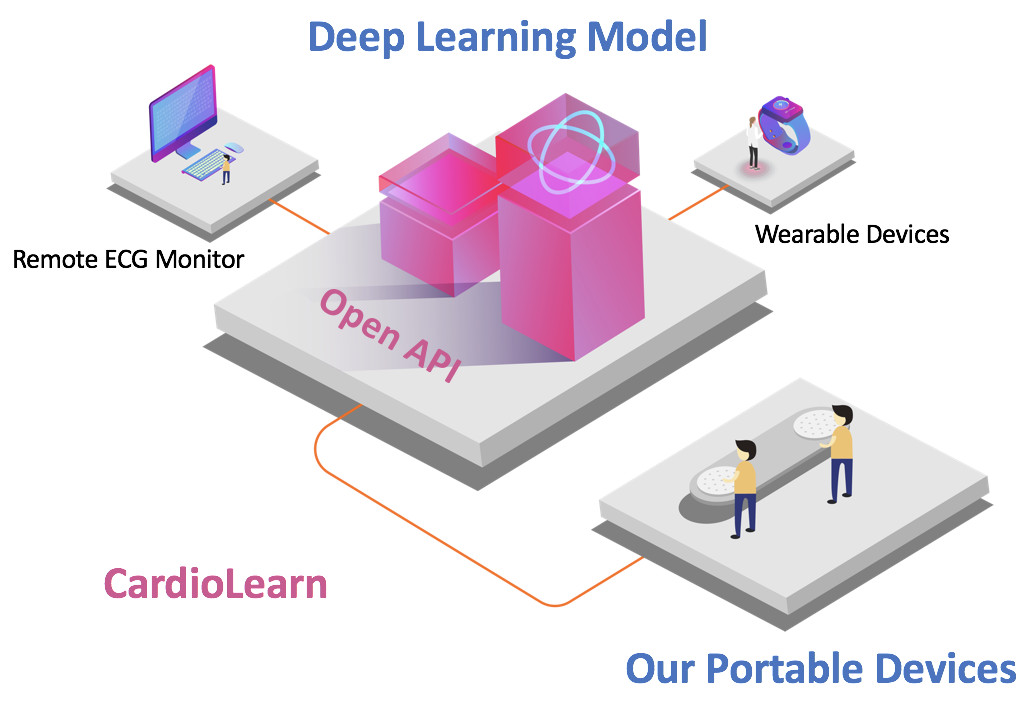}
\caption{The workflow of \mname ECG disease detection cloud deep learning service. }
\label{fig:framework}
\end{figure}

\section{System Design}

This section introduces our system in detail. We first introduce the details of our deep learning model and the performance tested on a publicly available open-source dataset. Then we show how we deploy the model and serve it as a cloud service. 

The framework of \mname is shown in Figure \ref{fig:framework}. We first build and train two deep neural network models to support applications in the healthcare environment (outside the hospital, usually single lead) and medical environment (inside the hospital, usually 12-lead). We are then serving the model by providing an open API using the HTTP protocol. Moreover, we also design a portable hardware device along with an interactive mobile program as a practical application. 

\subsection{A Deep Learning Model for ECG Disease Detection}

\begin{figure}[ht]
\includegraphics[width=0.48\textwidth]{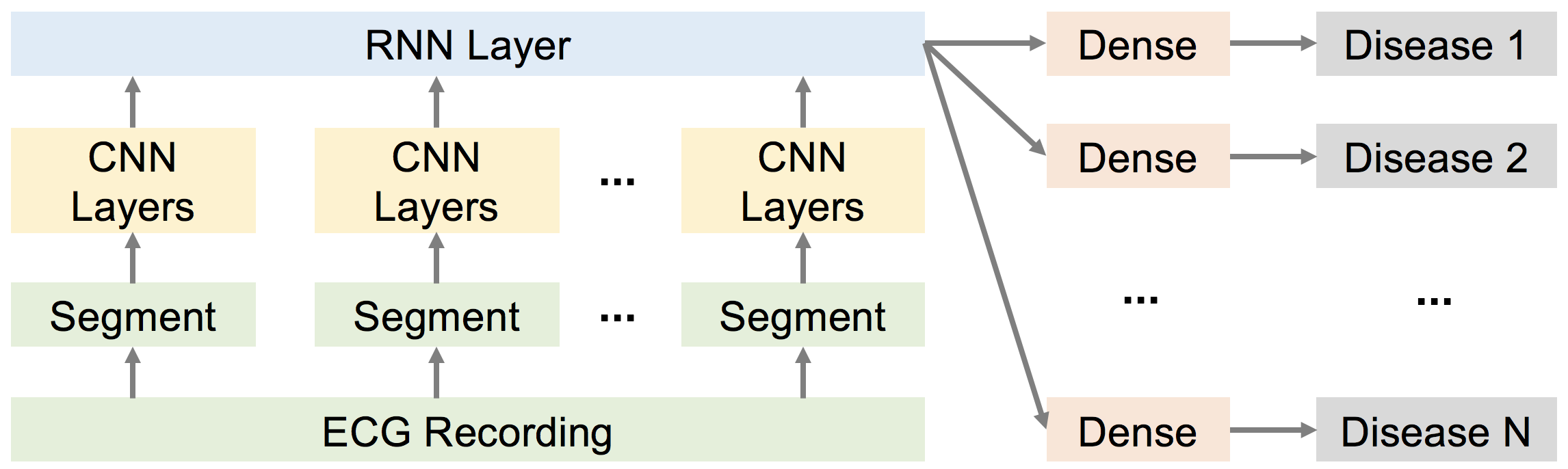}
\caption{Model architecture. }
\label{fig:model}
\end{figure}

\begin{figure}
\includegraphics[width=0.45\textwidth]{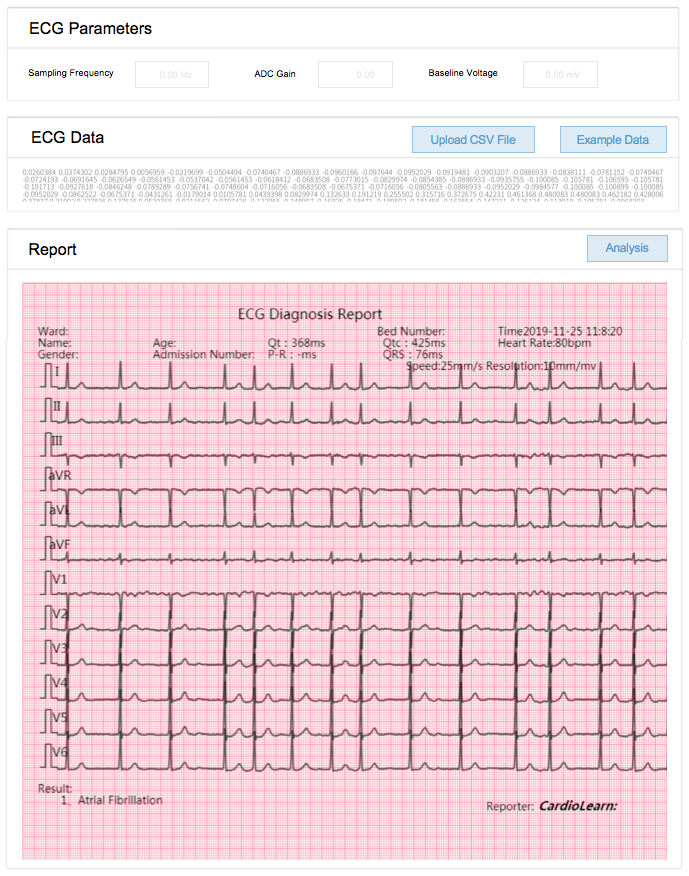}
\caption{Demo of Webpage for Cloud Deep Learning Service (12-lead ECG). }
\label{fig:demo2}
\end{figure}

\begin{figure*}
\includegraphics[width=0.95\textwidth]{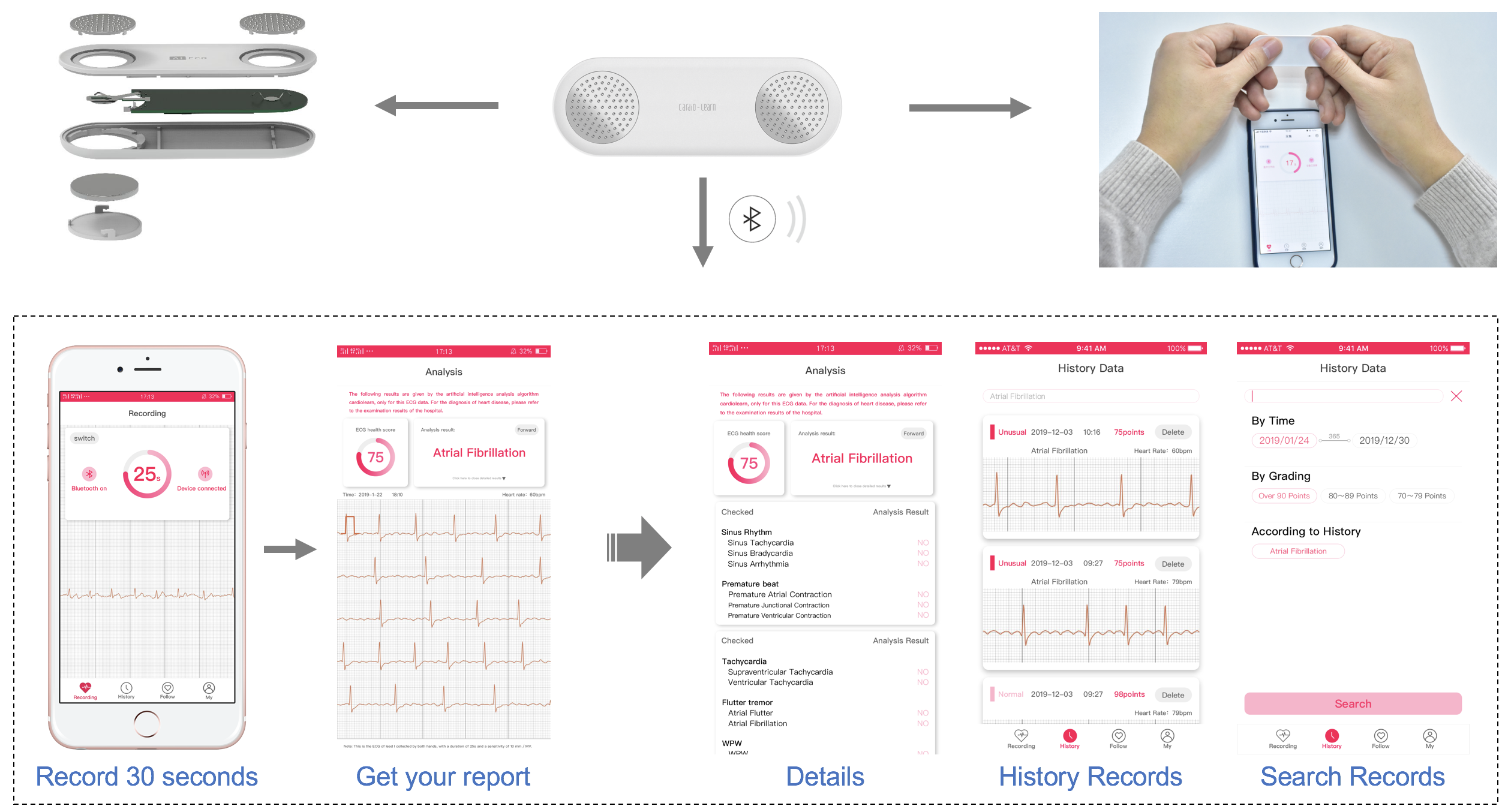}
\caption{Demo of Portable Hardware Device and Mobile Application (Single Lead ECG). }
\label{fig:demo1}
\end{figure*}

The ECG data usually has two forms of inputs, which are single lead (inside the hospital, see Figure \ref{fig:demo2}) and 12-lead (outside the hospital, see Figure \ref{fig:demo1}). Here ``lead'' has the same meaning as ``channel''. To handle them both, we build two deep neural network models for each kind using TensorFlow. Their input layers are different, while the other layers remain the same. In detail, as shown in Figure \ref{fig:model}, the input ECG recording is segmented into several short segments, and each segment goes through 32-layers of stacked one-dimensional convolutional layers (CNN) to capture local ECG patterns and shifts. One recurrent layer (RNN) is then built on top of the convolutional layers to capture long term variations. Finally, the model applies multiple dense layers (Dense) on the output of the recurrent layer to get the predictions of each disease. The objective of the model is a multi-label learning task because multiple diseases might occur in the same ECG recording. Moreover, we also introduce shortcut connections \cite{he2016identity} at every two convolutional layers to address the problems of vanishing/exploding gradients when training a very deep neural network. The input dimension is downsampled at every four convolutional layers, and the number of filters increases at every eight convolutional layers. The CNNs have shared weights between segments.

To train the deep model, we collected the training data from several hospitals, which are 12-lead ECG recordings lasting from 20 s to several minutes. The corresponding diagnosis results were written by cardiologists using narrative language; we extract keywords and integrate them into diagnostic labels based on standard ECG disease detection systems like \cite{mason2007recommendations}. We use Lead I as single lead data while also collect extra single lead data from mobile devices. Finally, our 12-lead model can support 43 types of diseases, and a single-lead model can support 18 types of diseases, both covering over 99\% of total abnormalities in our training data. 
We optimze the loss function by reducing the learning rate by a factor of 0.1 when the validation performance is not improving for 5,000 batches of each task. We are continuously saving and updating the best model for each label as the final model.

\begin{figure}[ht]
\begin{tabular}{cc}
\includegraphics[width=0.23\textwidth]{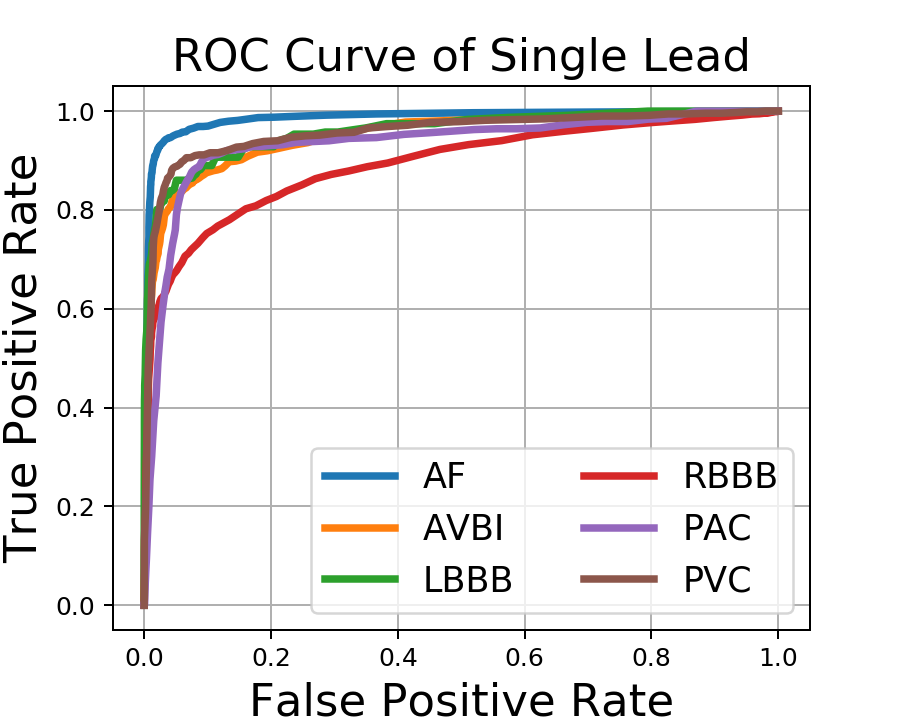}
& \includegraphics[width=0.23\textwidth]{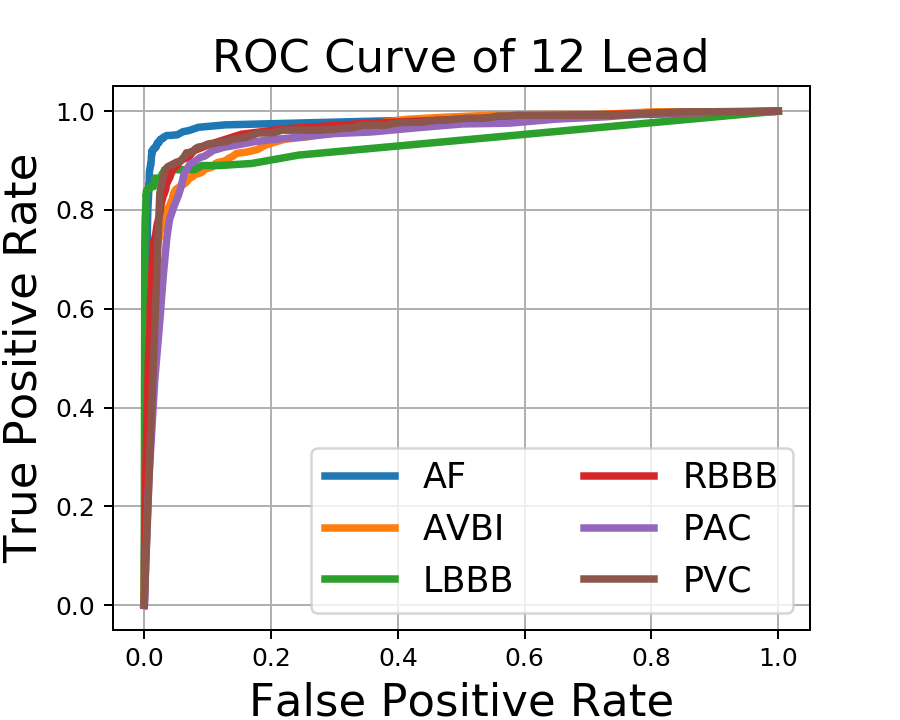}\\
\end{tabular}
\caption{ROC curves on 2018 China Physiological Signal Challenge dataset. }
\label{fig:roc}
\end{figure}

We tested our model on a publicly available open-source dataset from 2018 China Physiological Signal Challenge \footnote{\url{http://2018.icbeb.org/Challenge.html}} \cite{liu2018open}. The challenge ECG recordings were collected from 11 hospitals sampled as 500 Hz, which contains 6,877 (3178 female, 3699 male) 12-lead ECG recordings lasting from 6 s to just 60 s. These recordings are \textit{never} being used to train our model. We test the model performance on detecting Atrial fibrillation (AF), First-degree atrioventricular block (AVBI), Left bundle branch block (LBBB), Right bundle branch block (RBBB), Premature atrial contraction (PAC) and Premature ventricular contraction (PVC). We report the Receiver Operating Characteristic curve (ROC curve) and the area under ROC (ROC-AUC score) for each disease. The results are shown in Table \ref{tb:roc} and Figure \ref{fig:roc}. We can see that both models achieve higher than 0.93 ROC-AUC scores on almost all diseases. We also notice that AF detection is even higher than 0.97.

\begin{table}[ht]
\resizebox{1.0\linewidth}{!}{
\begin{tabular}{l|cccccc}
\toprule
            & AF     & AVBI   & LBBB   & RBBB   & PAC    & PVC    \\
\midrule
Single Lead & 0.9857 & 0.9508 & 0.9597 & 0.8927 & 0.9343 & 0.9578 \\
12-lead     & 0.9789 & 0.9579 & 0.9385 & 0.9655 & 0.9462 & 0.9609 \\
\bottomrule
\end{tabular}
}
\caption{ROC-AUC scores on 2018 China Physiological Signal Challenge dataset. }
\label{tb:roc}
\end{table}

\subsection{Serving Deep Model as a Cloud Engine}

We deploy our models using TensorFlow Serving \cite{olston2017tensorflow} on four cloud servers. 
Specifically, we use the Java Client TensorFlow Serving gRPC API. Each cloud server is equipped with 4-core Intel Xeon Skylake 6146 3.2 GHz CPU 16GB RAM, and 3 Tesla P4 GPU for model inference. The information transmission between servers and clients is based on HTTP protocol. One request of HTTP including HEADER and POST. The HEADER includes \textit{content type} like ``JSON'' and \textit{authorization} information. The authority is a unique token given by the server. The POST is \textit{content type} format includes \textit{sampleRate} (HZ, sample frequence of ECG signal), \textit{adcGain} (Analog-to-Digital Converter gain), \textit{dataI}, \textit{dataII}, \textit{dataIII}, \textit{dataAVR}, \textit{dataAVL}, \textit{dataAVF}, \textit{dataV1}, \textit{dataV2}, \textit{dataV3}, \textit{dataV4}, \textit{dataV5}, \textit{dataV6} represents 12 standard leads. One can only fill in \textit{dataI} and leave others null for requesting single lead model. The returned result is also in JSON format. One can easily parse the JSON result and integrate to their own systems.

The master server maintains a global task queue $Q$, which consists of all requests from clients. The task queue implements the First In First Out (FIFO) order. Each request contains authorization token $T$, analysis parameters $P$ (sampling rate, for example), and ECG data $D$. The disease detection process includes three steps: (1) authorization, (2) preprocessing, and (3) invoking deep model. Authorization validates the legality of the request by checking its unique token $T$. Once authorization success, the requests are added into the $Q$ and wait for the computation resources. Then, the engine resamples the original ECG and removes high-frequency noise as well as low frequency wandering by band-pass filters. After that, the engine invokes GPU and inference to get the results. If any step raises exception due to some errors, the request is added into task queue $Q$ again and wait for the next round analysis. If the process fails or timeout, the server returns a failure execution information. Notice that models are served by multiple concurrent processes so that the time delay of retry is expected to be short.

\begin{figure}
\includegraphics[width=0.45\textwidth]{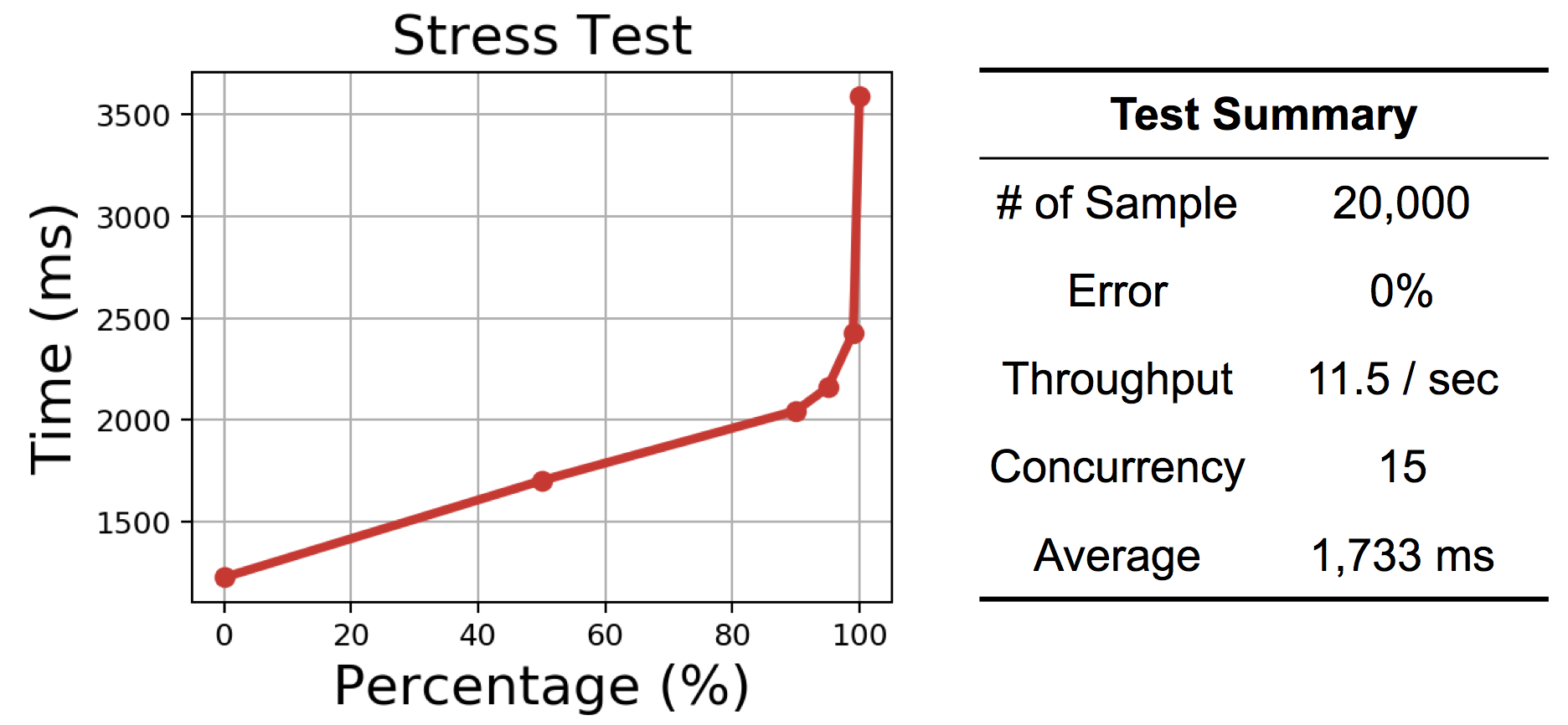}
\caption{Results of stress test. }
\label{fig:stress}
\end{figure}

We run the stress test to validate the performance of cloud serving using Apache Jmeter \footnote{\url{https://jmeter.apache.org}}. We prepare 20,000 samples 12-lead 30 seconds ECG recordings. The number of concurrent processes of each server is set to 15. The results are shown in Figure \ref{fig:stress}. The left figure shows the distribution of process time, and we can see that almost 99\% of requests can be returned within 2.5 seconds. This time cost shows a promising real-world application because it reduces one report of ECG disease detection from minutes (by Cardiologists) to seconds. The right figure shows the test summary. We can see that the total throughput is 11.5 records per second, which means \mname can analyze nearly 1 million 30-second ECG recordings per day. Notice that when handing long term ECG recordings, we can still keep a comparable execution time, by cutting long term ECG recordings into short recordings and batching them, where each batch can contain hundreds of short recordings. 

\section{Demonstration}

Our demonstration consists of two parts: 1) a website of an ECG analytic tool that provides an out-of-the-box cloud deep learning service, and 2) a portable hardware device to record ECG as well as an interactive mobile application to display results \footnote{For more information, please visit \url{https://github.com/hsd1503/CardioLearn}.}. 

Figure \ref{fig:demo2} shows the webpage of the ECG analytics tool. It consists of three steps to get analysis results. First of all, users should provide necessary parameters of their ECG data, including sampling frequency, ADC Gain, and baseline voltage. Then, users have two ways to upload their ECG data: 1) upload comma-separated values (CSV) file from their computer or 2) copy and paste CSV file into the website textbox directly. Besides, we also provide a variety of ECG records as example data. Finally, after clicking ``Analysis'' bottom, the formatted ECG report, including ECG records and analysis results, shows on the bottom as in Figure \ref{fig:demo2}. In the ECG report, the middle main part shows ECG recordings, with pink mesh grid background that helps cardiologists to measure and review the reports. The left bottom shows disease detection results given by \mname, which is Atrial Fibrillation in this case. The upper part shows ECG measurements like PR interval, QRS width, et al., which also help cardiologists for a better review.

Figure \ref{fig:demo1} shows our portable hardware device and interactive mobile application. 
The portable hardware device weighs around 10 g, and the size is 75 mm length, 25 mm width, and 4.5 mm height, which is very convenient to carry on in daily life. A Bluetooth module is equipped on the chip for connecting with a mobile phone. 
The interactive mobile application is developed based on WeChat Mini Program so that it can support any mobile phone, including Android or iOS, if they can install WeChat. The usage is simplified to only one step: just put the user's fingers on the metal electrodes and wait for 30 seconds. The device records ECG and sends the data to the application via Bluetooth transmission, the mobile application then sending the request to \mname for analysis, and finally display the returned results on the user's mobile phone. The ECG records and results can also be retrospected from the mobile application.

\section{Conclusion}

In this paper, we introduce \mname, a publicly available out-of-the-box cloud deep learning service for cardiac disease detection from ECG, which can help improve the analytic ability of existing ECG recording devices. Besides, we also design a portable smart hardware device along with an interactive mobile program to demonstrate such practical usage. We wish everyone can easily and early detect potential cardiac diseases anytime and anywhere.

\bibliographystyle{ACM-Reference-Format}
\bibliography{sample}

\end{document}